# Making Sense of How Students Make Sense of Mechanical Waves

Michael C. Wittmann, Richard N. Steinberg, Edward F. Redish
Dept. of Physics, University of Maryland, College Park 20742-4111

In our classroom experiences as teachers, we are often baffled when students correctly answer questions in one setting and then can't answer seemingly identical questions in another. Obviously, their understanding of the material is not as strong as we would like. But are we asking the relevant questions when we come to this conclusion? Do the students fundamentally not know the material? Do they know it but not recognize appropriate circumstances in which to use it? And how should our instruction and evaluation of their knowledge depend on the answers to these questions?

We have begun to address these questions at the University of Maryland using the methods and tools of physics education research.[1] Our approach combines the study of student difficulties with physics with the design of instructional materials and environments that help students improve their understanding. This approach can lead to educational environments that help students overcome their difficulties.[2]

We report here on our study of student understanding of the physics of mechanical waves. Understanding wave physics is important for making sense of physical optics, quantum mechanics, and electromagnetic radiation. Previous research has shown that students have fundamental difficulties with some of the basic concepts of wave physics.[1]

## Sense Making and Mental Models

To describe and understand our observations of student reasoning we need a framework and a language. In the University of Maryland Physics Education Research Group, we use the idea of *mental models* borrowed from the cognitive science and learning theory literature in order to interpret how students make sense of physics.[3,4] We use the term to mean the patterns of association and physical analogies that guide spontaneous responses and reasoning in unfamiliar situations. Mental models are less specific and less rigorous than physical models and are often not explicitly verbalized or consciously used.

For example, we have found that many students answering questions dealing with the propagation speed and superposition of mechanical waves appear to use a guiding analogy of waves that is reasonably complete and coherent, but at odds with physical reality. We call this the *Particle Pulses Mental Model*. We observe that many students do not think of the forces and interactions internal to a mechanical waves system as we do in the physical model of waves when we derive the wave equation or discuss its physical meaning. Instead, many students mistakenly use analogies with Newtonian particle mechanics and ideas of force, energy, and collisions between objects to describe the physics of waves. These analogies play an important role in how students come to an understanding of wave physics and how we test and probe their understanding in our classrooms.

## Research Context and Methods

Our investigations were done in the second semester of a three-semester university physics course at the University of Maryland. The course includes three hours of lecture a week, a traditional laboratory, and a small group discussion. Discussion sections are either traditional TA-led recitations or *tutorials*, a group-learning method developed by the Physics Education Group at the University of Washington.[5]

We use a variety of methods to investigate student understanding of physics. The work in this paper is part of an iterative cycle of research, curriculum development, delivery, and evaluation. We do our evaluation using a variety of probes, including videotaped individual demonstration interviews, pre-tests (short, ungraded quizzes that accompany tutorials), examination questions, and specially designed diagnostic tests. In interviews, individual students answer questions about a simple physical problem in a context where the researcher has the opportunity to probe their responses more deeply. Students who volunteer for these interviews are typically doing well in the class.

## Student Difficulties with Mechanical Waves

When beginning their study of waves, students need to learn two critical physical concepts of the small amplitude model of mechanical waves in a non-dispersive, deformable medium:

- Wave propagation is a medium's response to a disturbance. In the chosen model, the propagation characteristics depend only on the medium and not on the nature of the disturbance.
- Wave superposition occurs by adding individual displacements point by point at any given time.

Our observations indicate that many students have difficulties with both of these ideas, and their difficulties can often be described by a single mental model.

### Wave Propagation

Figure 1 shows a question about wave propagation speed in two formats: a free response question, and a multiple-choice, multiple-response (MCMR) question. The distractors in the MCMR question are based on answers students commonly give to free response and interview questions. Research-based distractors in multiple-choice problems have been used previously, but research has shown that this may only probe one aspect of student thinking.[6] In order to get a richer response, we give both free response and MCMR questions. In both formats, we ask students to explain their reasoning. We consider responses correct that state that only the medium properties (tension and string mass density) will affect the speed (answers 'e' and 'g' in the MCMR question).

After students had all instruction on waves in the Spring-1996 and Spring-1997 semesters, we asked the free response question in an interview setting.

We find that students use ideas of force and energy incorrectly in their reasoning. Some students use reasoning based on the force exerted by the hand to create the pulse. One student stated, "You flick [your hand] harder...you put a greater force in your hand, so it goes faster." Some students state that creating a wave with a larger amplitude takes greater force and thus the wave will move faster. Others state that shaking your hand harder (in interviews, this was usually accompanied by a quick jerk of the hand) will "put more force in the wave." Another student used reasoning based on energy to describe the effect of a change in hand motion. He stated, "We could make the initial pulse fast, if you flick [your hand], you flick it faster... It would put more

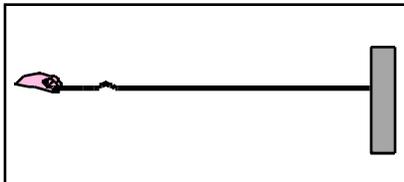

*Version 1: Free-Response format*: A taut string is attached to a distant wall. A pulse moving on the string towards the wall reaches the wall in time $t_0$ (see diagram). How would you decrease the time it takes for the pulse to reach the wall? Explain your

*Version 2: Multiple-Choice, Multiple-Response (MCMR) format*:
A taut string is attached to a distant wall. A demonstrator moves her hand to create a pulse traveling toward the wall (see diagram). The demonstrator wants to produce a pulse that takes a <u>longer time</u> to reach the wall. Which of the actions *a–k* **taken by itself** will produce this result? More than one answer may be correct. If so, give them all. Explain your reasoning.
a. Move her hand more quickly (but still only up and down once by the same amount).
b. Move her hand more slowly (but still only up and down once by the same amount).
c. Move her hand a larger distance but up and down in the same amount of time.
d. Move her hand a smaller distance but up and down in the same amount of time.
e. Use a heavier string of the same length, under the same tension
f. Use a lighter string of the same length, under the same tension
g. Use a string of the same density, but decrease the tension.
h. Use a string of the same density, but increase the tension.
i. Put more force into the wave.
j. Put less force into the wave.
k. None of the above answers will cause the desired effect.

**Figure 1: Wave Propagation Question in both Free Response and Multiple-Choice, Multiple-Response formats. Students answered both formats on diagnostic tests.**

energy in." This student is failing to distinguish between the velocity of the hand, which is associated with the transverse velocity of the string, and the longitudinal velocity of the pulse along the string.

Students using energy arguments to describe how they would change the speed of the wavepulse also often state that pulses of different sizes will move with different speeds. One student stated, "Make it wider, so that it covers more area, which will make it go faster." In follow-up comments, this student explained that it took more energy to create a larger pulse, and that the pulse would move faster because it had more energy. We have also found that some students state that <u>smaller</u> pulses will move faster. "Tinier, tighter hand movements" will allow the wave to slip more easily (thus, faster) through the medium.

In the Fall-1997 semester, we asked both question formats on diagnostic tests at the beginning of the semester before all instruction and near the end of the semester after all instruction on waves had been completed. In the beginning of the semester, students first answered the free response question, turned it in, and were then handed the MCMR question. This ensured that they did not change their answers on the free response question as a result of seeing the list of MCMR options. During the semester, instruction consisted of lecture, textbook homework problems, and tutorials designed to address the issues discussed in this paper. Based on preliminary research into student difficulties with wave propagation, we designed a set of video-based tutorials that help students visualize propagating waves.[7] These tutorials are designed to elicit student difficulties, confront students with incorrect predictions, and help students resolve any difficulties they have.[2b] After all instruction, students answered the free response and MCMR questions in successive weeks as a supplement to their weekly pretests given during lecture.

By comparing student responses on the two question formats, we can probe the distribution of ideas used by students to understand the physics of wave speed and the consistency of student reasoning. Table I shows how students answered both the free response and MCMR questions before instruction. Only those students who answered both types of questions both before and after instruction are included. Written explanations echo those given during interviews.

At the beginning of the semester, very few students give *only* the correct answer, but most of them include it in their responses. Almost all of the students answer that the hand motion will affect the wave speed. The results imply that students do not think consistently about the physics of wave speed.

Students predominantly use only one explanation when answering the free response question. The offered responses on the MCMR question appear to act as triggers that elicit additional explanations, especially from students who give the hand motion response on the free response question. Of the few students who answered the free response question using only correct reasoning, most answered the MCMR question consistently. These students seem to have a robust understanding of the dependence of wave speed on medium properties. More than three-fourths of the students emphasize the incorrect hand motion response at the beginning of the semester, though.

Table II shows student responses after they completed all instruction on waves. Student performance is somewhat improved, with more students giving completely correct explanations. Nearly all students recognize the correct answer on the MCMR question and nearly three-fourths give it for the free response question, but a majority of the class still believes that changes in hand motion play a role.

|  |  | Student responses on free response problem | | | |
|---|---|---|---|---|---|
|  | *Speed changes due to change in:* | only tension and density | both the medium and hand motion | The motion of the hand | other |
| Student responses on MCMR question | only tension and density | 7% | 1% | 2% | 1% |
|  | both the medium and hand motion | 1% | 2% | 60% | 10% |
|  | the motion of the hand | 1% | 1% | 11% | 3% |

**Table I: Comparison of student pre-instruction responses on free response and MCMR wave propagation questions, Fall-1997 (matched data, N=92). Students answered questions before all instruction.**

|  | Student responses on free response problem | | | |
|---|---|---|---|---|
| *Speed changes due to change in:* | only tension and density | both the medium and hand motion | the motion of the hand | other |
| **Student responses on MCMR question** — only tension and density | 40% | 2% | 2% | 2% |
| both the medium and hand motion | 8% | 17% | 20% | 2% |
| the motion of the hand | 2% | 1% | 2% | 0% |

**Table II: Comparison of student post-instruction responses on free response and MCMR wave propagation questions, Fall-1997 (matched data, N=92). Students answered questions after all instruction on waves.**

In both the pre and post instruction tables, the most common off-diagonal elements of the tables show that students who answer the free response question using only hand motion explanations are triggered into additionally giving correct medium change responses on the MCMR question. Apparently, they recognize the correct answer, but do not recall it on their own in a free response question. Fewer students are triggered in the other direction from correct medium change explanations on the free response question to additionally giving the hand motion response on the MCMR question. It is noteworthy that so many of the students answer incorrectly even after explicit instruction on the topic.

A summary of the matched MCMR data from Fall-1997 is shown in Figure 2. To provide a comparison to student performance without specially designed tutorials, we include (unmatched) data from a MCMR question asked in the Spring-1996 semester (no free response question was asked then). These results illustrate the contrast between pre-instruction, post-traditional instruction, and post-modified instruction student answers. While there is some improvement in student responses, the strength of the incorrect hand motion response is apparently quite robust.

### *Wave Superposition*

We have also investigated student understanding of superposition in the same environment.[8] Figure 3 shows a question we asked of 65 students in a pretest in the Spring-1996 semester.[9] In this situation, students had traditional lecture instruction on wave superposition before taking the pretest. A correct response to the question shows addition of displacements due to each wavepulse on a point-by-point basis. The correct response (with the individual wavepulse shapes indicated) and the two most common incorrect responses are shown in Figure 4.

From explanations given by students on the pretest and in follow-up interviews, we find that most students do not demonstrate a functional understanding of superposition (only 5% sketched Figure 4(a)). The most common incorrect response, given by 40% of the students, showed no superposition unless the peaks of the pulses overlapped (see Figure 4(b)). A typical student explanation was that "the waves only add when the amplitudes meet." We have found that students giving this explanation use the word "amplitude" to describe only the point of maximum displacement, and they ignore all other displaced points in their descriptions.

Other students also had difficulty with the process of wave addition. One-fifth of them sketched

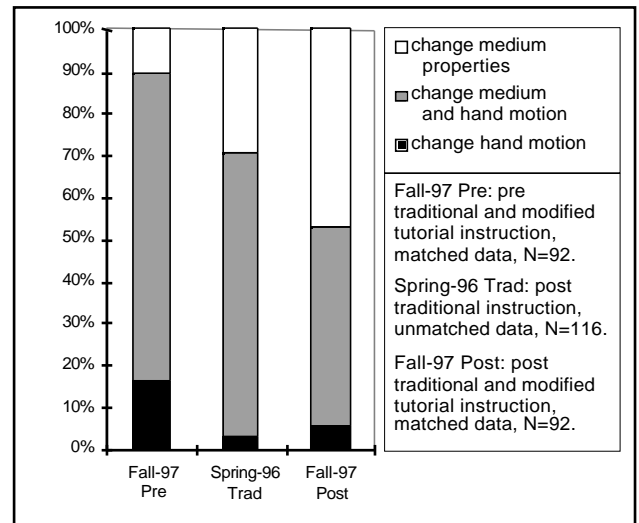

**Figure 2: Comparison of student responses on the MCMR wave propagation question, Fall-1997 (matched pre/post tutorial instruction, N=92) and Spring-1996 (unmatched, post traditional instruction, N=116). Students answered the question on diagnostic tests given before and after all instruction on waves.**

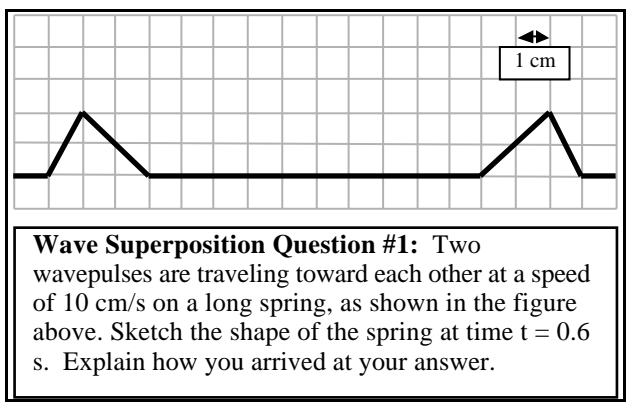

**Figure 3:** Wave superposition question from pretest given after traditional instruction, Spring-1996, N= 65. Students had completed lecture instruction on superposition.

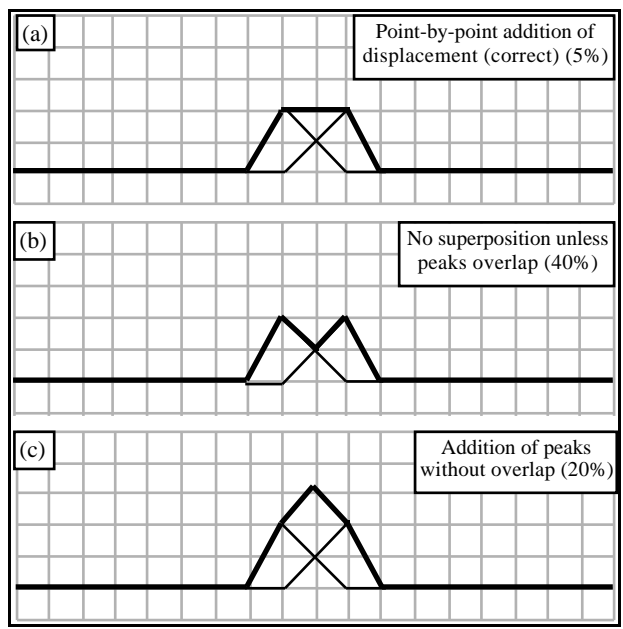

**Figure 4:** Common responses to pretest question from Spring-1996.
**(a) Correct response,**
**(b) common incorrect response,**
**(c) common incorrect response.**
**Individual wavepulse shapes are also shown. Student responses were given on pretests and in interviews which followed lecture instruction on superposition and preceded tutorial instruction.**

Figure 4(c) and stated that the points of maximum displacement would add even though they weren't at the same location on the string. This question was also asked in an interview setting. One interviewed student who used the word "amplitude" incorrectly, as described above, explained, "Because the [bases of the] waves are on top of another, the amplitudes add."

Students giving the described incorrect responses seem to view superposition as the addition of the maximum displacement point only and not as the addition of displacement at <u>all</u> locations. They do not have a complete understanding of wavepulses as extended regions that are displaced from equilibrium. Instead, they seem to simplify the wavepulse to a single point.

At the beginning of the Fall-1995 semester, before all instruction, we asked 182 students a diagnostic test that included the question shown in Figure 5. A correct response to the question (shown in Figure 6(a)), given by 55% of the students, shows that the wavepulses pass through each other. Superposition of waves should have no permanent effect on the two waves.

The most common incorrect answer, shown in Figure 6(b), was given by one fifth of the students. One student wrote, "[Part of] the greater wave is canceled by the smaller one." In explanations, students implied that they were thinking of this as a collision. If we consider two carts of unequal size moving toward each other at the same speed and colliding in a perfectly inelastic collision (imagine Velcro$^{TM}$ holding them together), then the unit of two carts would continue to move in the direction the larger was originally moving, but at a slower speed.

One student's written comment supports this interpretation, "The smaller wave would move to the right, but at a slower speed." These students appear to be thinking of wavepulses as objects that collide with each other or cancel one another out. In other questions, students sometimes state that same-size pulses bounce off each other.

### *The Particle Pulses Mental Model of Mechanical Waves*

To describe student difficulties with waves, we organize the data in terms of a *mental model*. As described earlier, mental models are the patterns of associations (i.e. rules, images, maps, or analogies) used to guide spontaneous reasoning.[3] But student mental models are often incomplete, self-contradictory, and inconsistent with experimental data.

When trying to organize student responses to questions described in the previous section, we see evidence for what we call the *Particle Pulses Mental Model* of waves. Students using this mental model

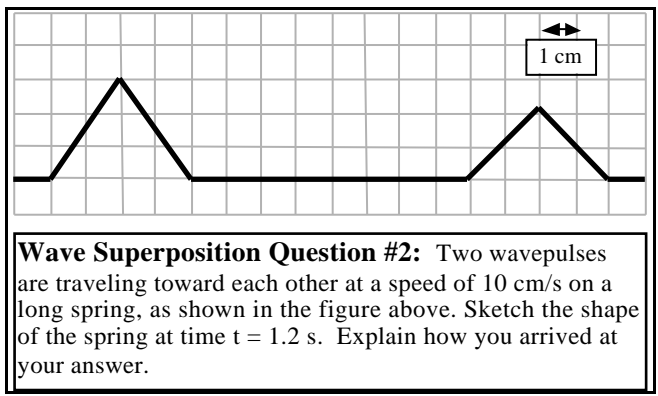

**Wave Superposition Question #2:** Two wavepulses are traveling toward each other at a speed of 10 cm/s on a long spring, as shown in the figure above. Sketch the shape of the spring at time t = 1.2 s. Explain how you arrived at your answer.

Figure 5: Wave superposition question from a diagnostic test given, F95 semester (N= 182 students). Students had no instruction on waves when they took this diagnostic.

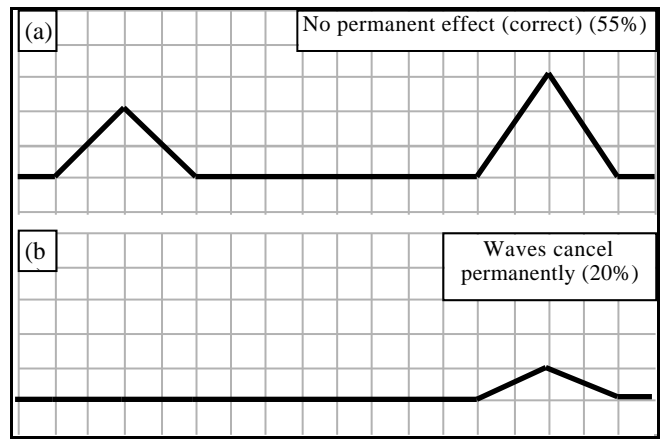

Figure 6: Common responses to diagnostic question from F95. (a) Correct response, (b) Most common incorrect response.

typically make analogies to mechanical particle-physics models. The Particle Pulses Mental Model involves the incorrect use of force or energy arguments and an inability to look at local characteristics of the wave. Though some parts of the Particle Pulses Mental Model are reminiscent of student common sense beliefs of mechanics,[10] others are unique to waves. Table III gives a summary of our description. We find that some students have a superposition of mental models with a distribution that depends on their understanding and the "trigger" of the question presented.

Students who use the Particle Pulses Mental Model seem to make an analogy between wave propagation and throwing a rigid object like a ball. If one throws a ball harder, it goes faster. Many students seem to use this analogy to guide their reasoning about wave speed; greater force in the hand motion creates a greater speed. In interviews, we have heard many students state that waves exert a force on the medium through which they travel and push the medium along like a surfer on a wave. The "surfer" description is common when students describe how sound waves affect air. In this context, sinusoidal waves are often described as a succession of pulses, each exerting a force (or "kick") in the direction of wave propagation.

Other students seem to make the ball-toss analogy when using energy arguments to describe how to change wave speed. A ball with a larger kinetic energy whose mass remains constant moves faster. Similarly, a pulse with more energy whose size stays constant must move faster. The explanation that a smaller mass will move faster is consistent with this explanation, too, because a smaller mass, with energy held constant, has a larger velocity. Though students do not explicitly state the analogy between their descriptions of wave speed and a thrown ball, their descriptions in interviews are often consistent with a mental model built on this analogy.

Students often give point-particle descriptions of wavepulses. The ball-toss analogy gives an example of this reasoning in wave propagation. Similarly, in superposition, many students give the response that the wavepulses do not add until the points of maximum displacement overlap. They treat each wavepulse as a single point and ignore all other points. Other students describe the entire wavepulse as a single point, not ignoring the non-peak displaced points, but lumping all displaced points together into one.

Many students use a collision-like description of wave superposition to describe the interaction of wavepulses. Superposing wavepulses collide with each other and either bounce off each other or cancel out, depending on the situation. The remnant wavepulse possibly moves slower, having lost energy during the collision. Here we see a clear example of the way an incorrectly used mental model plays a role in student reasoning. We have not found that students will *explicitly* state that the waves act like colliding carts, but we find that they often give descriptions consistent with the physics of cart collisions. Because of the similar explanations for the two situations, we believe students have an associative pattern that guides their understanding of wave interactions.

| Newtonian Particle Model | Particle Pulses Mental Model | Simplified Wave Model |
|---|---|---|
| A harder throw implies a faster particle. | A harder flick of the wrist implies a faster wavepulse. | Wave speed depends only on medium response to disturbance. |
| Smaller objects are more easily thrown faster. | Smaller pulses can be created that move faster. | Size of pulse and manner of wave creation do not affect wave speed. |
| An object's center of mass is considered when describing its motion (trajectory). | Only the peak of the wavepulse is considered when describing superposition. | It is necessary to consider the entire shape of a wave to describe its properties (e.g. in superposition). |
| Objects collide with each other and their motion changes | Wavepulses collide with each other and they cancel or bounce off each other. | Waves pass through each other with no permanent effect. |

**Table III: Comparing student responses indicative of the Newtonian particle model, the Particle Pulses Mental Model, and the simplified wave physics model (in the small angle, non-dispersive medium approximation). Many students use elements of several models when answering questions concerning wave physics.**

## Implications for Assessment and Classroom Practice

We have found great value in using a variety of question formats to investigate student understanding of one topic. As physics education researchers, we have been able to carry out more than 40 hours of interviews that give evidence of both student strengths and weaknesses understanding the physics of waves. As instructors, we also would like to probe student understanding, but it is difficult to conduct large numbers of detailed interviews. In probing the development of mental models of mechanical waves, we have used both free response and MCMR questions to gain insight into student understanding that we would not otherwise have gained.

Answers to the free response question show where students are having continuing difficulty with the material. However, focusing only on difficulties gives an incomplete picture of student understanding. Answers to the MCMR question indicate that students use elements of many different mental models. This complements what is learned from the free response question. Using the two techniques together, we have been able to probe the manner in which students use multiple models in their reasoning about wave physics. Classroom instructors could find value in using both types of questions.

Often in physics, we expect our students to understand and apply well-defined, coherent models of physical systems. But our students often have a fragmented picture of physics. They seem to access their knowledge depending on criteria triggered by the question and situation at hand. Thus, they may simultaneously have both correct and incorrect ideas about specific physical situations. Both as instructors and as physics education researchers, we benefit from an understanding of the elements of students' reasoning and the criteria by which students organize their understanding.

## Acknowledgments

The authors would like to thank Jeff Saul and Mel Sabella for their contributions to the research discussed in this paper. The research in this paper is part of a dissertation by Michael Wittmann to be published by the University of Maryland. It has been funded in part by NSF grants DUE#-9455561 and DUE#-9652877.